\documentclass[4apaper,numberedheadings,draft]{aipproc}
\layoutstyle{8x11single}

\begin{document}

\title{Delirium Quantum\footnote{This piece was originally published  in G.~Adenier, C.~A. Fuchs, and A.~Yu.\ Khrennikov, editors, {\sl Foundations of Probability and Physics -- 4}, AIP Conference Proceedings Vol.~889, (American Institute of Physics, Melville, NY, 2007), pp.~438--462.  The author's present affiliation is: Perimeter Institute for Theoretical Physics, 31 Caroline Street North,
Waterloo, Ontario N2L 2Y5, Canada.} \medskip\\
\it Or, where I will take quantum mechanics if it will let me}

\author{Christopher A. Fuchs}{address = {Bell Labs, Lucent Technologies,
600-700 Mountain Avenue, Murray Hill, New Jersey 07974, USA}}

\date{14 December 2006}


\newtheorem{wiseman}{Wisemanism}
\newtheorem{sudbery}{Sudberyism}
\newtheorem{musser}{Musserism}

\newcommand{\bhw}{\begin{wiseman}\protect$\!\!${\em\bf :}$\;\;$}
\newcommand{\ehw}{\end{wiseman}}
\newcommand{\bts}{\begin{sudbery}\protect$\!\!${\em\bf :}$\;\;$}
\newcommand{\ets}{\end{sudbery}}
\newcommand{\bgm}{\begin{musser}\protect$\!\!${\em\bf :}$\;\;$}
\newcommand{\egm}{\end{musser}}

\newcommand{\bq}{\begin{quotation}}
\newcommand{\eq}{\end{quotation}}


\renewcommand{\thefootnote}{}

\begin{abstract}
Once again, I take advantage of the wonderfully liberal and tolerant
mood Andrei Khrennikov sets at his yearly conferences by submitting a
nonstandard paper for the proceedings. This pseudo-paper consists of
excerpts drawn from two of my samizdats [{\sl Quantum States:\ What
the Hell Are They?}\ and {\sl Darwinism All the Way Down (and
Probabilism All the Way Back Up)\/}\/] that I think best summarize
what I am aiming for on the broadest scale with my quantum
foundations program. Section 1 tries to draw a picture of a physical
world whose essence is ``Darwinism all the way down.'' Section 2
outlines how quantum theory should be viewed in light of that, i.e.,
as being an expression of probabilism (in Bruno de Finetti or Richard
Jeffrey's sense) all the way back up. Section 3 describes how the
idea of ``identical'' quantum measurement outcomes, though sounding
atomistic in character, nonetheless meshes well with a William
Jamesian style ``radical pluralism.'' Sections 4 and 5 further detail
how quantum theory should not be viewed so much as a ``theory of the
world,'' but rather as a theory of decision-making for agents {\it
immersed\/} within a quantum world---that is, a world in continual
creation. Finally, Sections 6 and 7 attempt to sketch once again the
{\it very positive\/} sense in which quantum theory is incomplete,
but still just as complete is it can be. In total, I hope these heady
speculations convey some of the excitement and potential I see for
the malleable world quantum mechanics hints of.
\end{abstract}



\maketitle

\section{to H.~M. Wiseman, 24 June 2002, ``The World is Under
Construction''}

\bhw
Do you believe that events in the world really are random? Or do
you believe they only appear to be random? In the first case,
doesn't that mean that you have to believe in objective
probabilities? \ldots

Or are you saying that the real world is unanalysable, unthinkable
even? Everything we say should be couched in terms of gambling
commitments. First, that seems to be a cop-out, giving up on any
understanding of the Universe. Second, it can't explain anything in
the Darwinian way you mentioned, except Dutch-book consistency. It
can't explain why it is ``bad'' to hold a gambling commitment based
on the idea that all world cup soccer balls contain bombs that have a
50\% chance to blow up every time a goal is scored \ldots\ \ You
cannot say anything about animals that would have been likely to have
gone extinct because of poor (but consistent) gambling commitments,
because that is a statement using the concept of objective
probabilities. You cannot\ {\bf\underline{explain}} anything that is
not strictly deterministic without using objective probabilities, it
seems to me.

I trust you understand my motives. I wouldn't bother discussing
this with you if I didn't think your ideas were potentially
revelational. What does not kill you makes you stronger.
\ehw

Of the three options you gave me for answering your questions, I
suppose if I were forced to choose one, I would align myself with the
one you called a ``cop-out.''  However, from my point of view, the
language you use builds about the ugliest picture it can for where
this effort is going. Indeed, you miss the very point, the very
beauty, of the ``cop-out.'' So, what I'd like to do is set that
right---right here and right now---before we go much further.

You see, the very starting point for most of my latest thoughts---the
thing I think quantum mechanics gives us the deepest and most
thorough hint of---is that there is no such thing as THE universe in
any completed and waiting-to-be-discovered sense.  The thought I am
\underline{\it testing out}\, is that the universe as a whole is
still under construction. And when I say this, I am not thinking of
just bits and pieces of it; I am thinking of the whole shebang, all
the way to the roots.  Nothing is completed.  Not just the playhouse
Kiki is building for Emma and Katie, or the evolutionary track of the
human species, but even the ``very laws'' of physics. The idea is
that they too are building up in precisely the way---and ever in the
same danger of falling down as---individual organic species. That is
to say, it's Darwinism all the way down.

So when you ask me if I am ``saying that the real world is
unanalysable, unthinkable even,'' the answer in a way is ``yes.''
For it is blatantly impossible to analyze to the last detail the
characteristics of a world that has not even been dreamt up (even
in its own mind's eye).

But how can I impress this upon you, or even make it seem reasonable
as a direction for research?  That is a tough call. For, like with
beer or single-malt Scotch, it is surely an acquired taste that
builds only slowly and with the right company. Of course, I could
just send you back to my paper quant-ph/0204146, ``The Anti-V\"axj\"o
Interpretation of Quantum Mechanics,''~\cite{Fuchs02} and ask you to
take it very seriously. But this morning it dawned on me to maybe
spend a little time with my scanner to try to ``IV'' some thoughts
straight into your bloodstream.

At the moment, I can think of no better introductions to the line of
thought I'd like to expose you to than three articles by Richard
Rorty:  ``A World without Substances or Essences,'' ``Truth without
Correspondence to Reality,'' and ``Thomas Kuhn, Rocks, and the Laws
of Physics.''$^1$\footnote{
\bq
\noindent $^1$WARNING: \ Just because I say I can think of no better
introductions to these ideas, it does not mean I endorse every
statement in these papers; I may not endorse half of them. However, I
think these papers go in the right direction, even if they go too far
\ldots\ and even if their arguments are far too weak. But I choose
the papers I do because they are easy reading, with beautiful
writing, and I suspect these thoughts are so foreign to you that if
you can find any sense in {\it some of them}, then it may be a good
start for a dialogue. Moreover, I continue to stress that the best
justification yet to pursue this direction of thought---and this is
something Rorty does not know---is quantum mechanics itself.  So,
rather than being the final words on things, these are just the
beginning words on things.
\eq
} (Read them in that order, if you read them.) All three papers can
be found in his collection of essays, {\sl Philosophy and Social
Hope}~\cite{Rorty99}. If you absorb these, I think you'll understand
completely what I'm up to, and why I so dislike the negative
connotations you associate with the radical-Bayesian way of viewing
the quantum state. Of course, it may not turn your head the way it
turns mine, but at least you'll know where I'm coming from, and from
what pool of enthusiasm I derive my strength to eschew the ``golden
nuggets'' of {\it mere\/} quantum cosmology, {\it mere\/} Bohmianism,
and {\it mere\/} ``dreams of a final theory.''  The world as I see it
is a much bigger place than those stories can tell.  And the
interpretational issues at the core of quantum mechanics strike me as
our first rigorous indication that there is something more to this
idea than simply the hopes and desires of an enthusiast.

For now, let me give you a flavor of the thoughts in these papers,
and then leave you on your own in the case that you would like to
pursue this further.  The following quotes come from ``Truth
without Correspondence to Reality.''

\bq
In this essay I shall focus on Whitman's phrase `counts \ldots\
for her justification and success \ldots\ almost entirely upon the
future'. As I see it, the link between Whitmanesque Americanism
and pragmatist philosophy---both classical and `neo-'---is a
willingness to refer all questions of ultimate justification to
the future, to the substance of things hoped for. If there is
anything distinctive about pragmatism it is that it substitutes
the notion of a better human future for the notions of `reality',
`reason' and `nature'. One may say of pragmatism what Novalis said
of Romanticism, that it is `the apotheosis of the future'.

As I read Dewey, what he somewhat awkwardly called `a new metaphysic
of man's relation to nature', was a generalization of the moral of
Darwinian biology. The only justification of a mutation, biological
or cultural, is its contribution to the existence of a more complex
and interesting species somewhere in the future.  Justification is
always justification from the point of view of the survivors, the
victors; there is no point of view more exalted than theirs to
assume. This is the truth in the ideas that might makes right and
that justice is the interest of the stronger. But these ideas are
misleading when they are construed metaphysically, as an assertion
that the present status quo, or the victorious side in some current
war, stand in some privileged relation to the way things really are.
So `metaphysic' was an unfortunate word to use in describing this
generalized Darwinism which is democracy. For that word is associated
with an attempt to replace appearance by reality.

Pragmatists---both classical and `neo-'---do not believe that there
is a way things really are. So they want to replace the
appearance-reality distinction by that between descriptions of the
world and of ourselves which are less useful and those which are more
useful. When the question `useful for what?'\ is pressed, they have
nothing to say except `useful to create a better future'. When they
are asked, `Better by what criterion?', they have no detailed answer,
any more than the first mammals could specify in what respects they
were better than the dying dinosaurs. Pragmatists can only say
something as vague as: Better in the sense of containing more of what
we consider good and less of what we consider bad. When asked, `And
what exactly do you consider good?', pragmatists can only say, with
Whitman, `variety and freedom', or, with Dewey, `growth'. `Growth
itself,' Dewey said, `is the only moral end.'

They are limited to such fuzzy and unhelpful answers because what
they hope is not that the future will conform to a plan, will
fulfil an immanent teleology, but rather that the future will
astonish and exhilarate. Just as fans of the avant garde go to art
galleries wanting to be astonished rather than hoping to have any
particular expectation fulfilled, so the finite and
anthropomorphic deity celebrated by James, and later by A. N.
Whitehead and Charles Hartshorne, hopes to be surprised and
delighted by the latest product of evolution, both biological and
cultural. Asking for pragmatism's blueprint of the future is like
asking Whitman to sketch what lies at the end of that illimitable
democratic vista. The vista, not the endpoint, matters.

So if Whitman and Dewey have anything interesting in common, it is
their principled and deliberate fuzziness. For principled
fuzziness is the American way of doing what Heidegger called
`getting beyond metaphysics'. As Heidegger uses it, `metaphysics'
is the search for something clear and distinct, something fully
present. That means something that does not trail off into an
indefinite future \ldots
\eq
and
\bq
So far I have been trying to give an overview of Dewey's place in
the intellectual scheme of things by saying something about his
relation to Emerson, Whitman, Kant, Hegel and Marx. Now I want to
become a bit more technical, and to offer an interpretation of the
most famous pragmatist doctrine---the pragmatist theory of truth.
I want to show how this doctrine fits into a more general
programme: that of replacing Greek and Kantian dualisms between
permanent structure and transitory content with the distinction
between the past and the future. I shall try to show how the
things which James and Dewey said about truth were a way of
replacing the task of justifying past custom and tradition by
reference to unchanging structure with the task of replacing an
unsatisfactory present with a more satisfactory future, thus
replacing certainty with hope. This replacement would, they
thought, amount to Americanizing philosophy. For they agreed with
Whitman that America is the country which counts for its `reason
and justification' upon the future, and {\it only\/} upon the
future.

Truth is what is supposed to distinguish knowledge from
well-grounded opinion---from justified belief. But if the true is,
as James said, `the name of whatever proves itself to be good in
the way of belief, and good, too, for definite, assignable,
reasons', then it is not clear in what respects a true belief is
supposed to differ from one which is merely justified. So
pragmatists are often said to confuse truth, which is absolute and
eternal, with justification, which is transitory because relative
to an audience.

Pragmatists have responded to this criticism in two principal ways.
Some, like Peirce, James and Putnam, have said that we can retain an
absolute sense of `true' by identifying it with `justification in the
ideal situation'---the situation which Peirce called `the end of
inquiry'. Others, like Dewey (and, I have argued, Davidson), have
suggested that there is little to be said about truth, and that
philosophers should explicitly and self-consciously {\it confine\/}
themselves to justification, to what Dewey called `warranted
assertibility'.

I prefer the latter strategy. Despite the efforts of Putnam and
Habermas to clarify the notion of `ideal epistemic situation',
that notion seems to me no more useful than that of
`correspondence to reality', or any of the other notions which
philosophers have used to provide an interesting gloss on the word
`true'. Furthermore, I think that any `absoluteness' which is
supposedly ensured by appeal to such notions is equally well
ensured if, with Davidson, we insist that human belief cannot
swing free of the nonhuman environment and that, as Davidson
insists, most of our beliefs (most of {\it anybody's\/} beliefs)
must be true. For this insistence gives us everything we wanted to
get from `realism' without invoking the slogan that `the real and
the true are ``independent of our beliefs''\,'---a slogan which,
Davidson rightly says, it is futile either to accept or to reject.

Davidson's claim that a truth theory for a natural language is
nothing more or less than an empirical explanation of the causal
relations which hold between features of the environment and the
holding true of sentences, seems to me all the guarantee we need
that we are, always and everywhere, `in touch with the world'. If
we have such a guarantee, then we have all the insurance we need
against `relativism' and `arbitrariness'. For Davidson tells us
that we can never be more arbitrary than the world lets us be. So
even if there is no Way the World Is, even if there is no such
thing as `the intrinsic nature of reality', there are still causal
pressures. These pressures will be described in different ways at
different times and for different purposes, but they are pressures
none the less.

The claim that `pragmatism is unable to account for the
absoluteness of truth' confuses two demands: the demand that we
explain the relation between the world and our claims to have true
beliefs and the specifically epistemological demand either for
present certainty or for a path guaranteed to lead to certainty,
if only in the infinitely distant future. The first demand is
traditionally met by saying that our beliefs are made true by the
world, and that they correspond to the way things are. Davidson
denies both claims. He and Dewey agree that we should give up the
idea that knowledge is an attempt to {\it represent\/} reality.
Rather, we should view inquiry as a way of using reality. So the
relation between our truth claims and the rest of the world is
causal rather than representational. It causes us to hold beliefs,
and we continue to hold the beliefs which prove to be reliable
guides to getting what we want. Goodman is right to say that there
is no one Way the World Is, and so no one way it is to be
accurately represented. But there are lots of ways to act so as to
realize human hopes of happiness. The attainment of such happiness
is not something distinct from the attainment of justified belief;
rather, the latter is a special case of the former.

Pragmatists realize that this way of thinking about knowledge and
truth makes certainty unlikely. But they think that the quest for
certainty---even as a long-term goal---is an attempt to escape
from the world. So they interpret the usual hostile reactions to
their treatment of truth as an expression of resentment,
resentment at being deprived of something which earlier
philosophers had mistakenly promised. Dewey urges that the quest
for certainty be replaced with the demand for imagination---that
philosophy should stop trying to provide reassurance and instead
encourage what Emerson called `self-reliance'. To encourage
self-reliance, in this sense, is to encourage the willingness to
turn one's back both on the past and on the attempt of `the
classical philosophy of Europe' to ground the past in the eternal.
It is to attempt Emersonian self-creation on a communal scale. To
say that one should replace knowledge by hope is to say much the
same thing: that one should stop worrying about whether what one
believes is well grounded and start worrying about whether one has
been imaginative enough to think up interesting alternatives to
one's present beliefs. As West says, `For Emerson, the goal of
activity is not simply domination, but also provocation; the telos
of movement and flux is not solely mastery, but also stimulation.'
\eq
and
\bq
It may seem strange to say that there is no connection between
justification and truth. This is because we are inclined to say
that truth is the aim of inquiry. But I think we pragmatists must
grasp the nettle and say that this claim is either empty or false.
Inquiry and justification have lots of mutual aims, but they do
not have an overarching aim called truth. Inquiry and
justification are activities we language-users cannot help
engaging in; we do not need a goal called `truth' to help us do
so, any more than our digestive organs need a goal called health
to set them to work. Language-users can no more help justifying
their beliefs and desires to one another than stomachs can help
grinding up foodstuff. The agenda for our digestive organs is set
by the particular foodstuffs being processed, and the agenda for
our justifying activity is provided by the diverse beliefs and
desires we encounter in our fellow language-users. There would
only be a `higher' aim of inquiry called `truth' if there were
such a thing as {\it ultimate\/} justification---justification
before God, or before the tribunal of reason, as opposed to any
merely finite human audience.

But, given a Darwinian picture of the world, there can be no such
tribunal. For such a tribunal would have to envisage all the
alternatives to a given belief, and know everything that was
relevant to criticism of every such alternative. Such a tribunal
would have to have what Putnam calls a `God's eye view'---a view
which took in not only every feature of the world as described in
a given set of terms, but that feature under every other possible
description as well. For if it did not, there would remain the
possibility that it was as fallible as the tribunal which sat in
judgment on Galileo, a tribunal which we condemn for having
required justification of new beliefs in old terms. If Darwin is
right, we can no more make sense of the idea of such a tribunal
than we can make sense of the idea that biological evolution has
an aim. Biological evolution produces ever new species, and
cultural evolution produces ever new audiences, but there is no
such thing as the species which evolution has in view, nor any
such thing as the `aim of inquiry'.

To sum up, my reply to the claim that pragmatists confuse truth
and justification is to turn this charge against those who make
it. They are the ones who are confused, because they think of
truth as something towards which we are moving, something we get
closer to the more justification we have. By contrast, pragmatists
think that there are a lot of detailed things to be said about
justification to any given audience, but nothing to be said about
justification in general. That is why there is nothing general to
be said about the nature or limits of human knowledge, nor
anything to be said about a connection between justification and
truth. There is nothing to be said on the latter subject not
because truth is atemporal and justification temporal, but because
{\it the {\bf only} point in contrasting the true with the merely
justified is to contrast a possible future with the actual
present}.
\eq

I don't have to tell you that I find these ideas tremendously
exciting.  It is not that nature is hidden from us.  It is that it is
not all there yet and never will be; `nature' is being hammered out
as we speak.  And just like with a good democracy, we all have a
nonnegligible input into giving it shape.  That is the idea I am
\underline{\it testing}\/ for consistency and utility.  On the chance
that it will lead somewhere, it seems to me, worth the gamble.

\section{to H.~M. Wiseman, 27 June 2002, ``Probabilism All the Way Up''}

\bhw
[Y]ou say that my language ``builds about the ugliest picture it can
for where this effort is going''. As I keep saying, I mean to be
provocative. I hope it drives you to new heights in building a
beautiful picture in response. Honestly I do see the beauty in your
program. And I think the more extreme it becomes, the more beautiful
it becomes. I am very interested to see where it ends up.
\ehw

Thanks for the compliment.  And, indeed, your correspondence does
drive me to new heights (of something).  But now I worry that I
offended you with my phrase ``ugliest picture.''  It probably came
off that way, but it wasn't meant to be an emotional statement or
a point about you personally.  If some emotion did slip into it,
it most likely refers to a conversation I had with Harvey Brown,
circa September 11 of last year.  Harvey kept saying that I wanted
to ``doom'' nature to being ``ineffable.''  But that language
carries such a negative connotation.  It carries the idea that
there is something there that we can never, or should never,
attempt to speak of.  So, when you said something similar in
print, it gave me the opportunity to try to reply in print.  (As
you know, I try to have my thoughts recorded so I can refer people
to them.  One of the original ideas was that it would save me time
that way; so far, that aspect of it hasn't worked out.)  Anyway,
as I made clear, I want to combat that with all my strength.  In
particular, the way that I am thinking about it, it is not a bad
thing that there are some things beyond description in nature.
Instead, it is just a statement that there are more things to
come; it is a way of leaving room for something new.

\bhw
[A]s it happens, I was reading a critique of Richard Rorty the very
morning before I got your letter. Otherwise I never would have heard
of him. It was a 1997 article by Alex Callinicos ``Postmodernism:\ a
critical diagnosis''~\cite{Callinicos97}. The most interesting
criticism in there was to say that Rorty ``presumes what he needs to
establish, namely that science and philosophy can be assimilated into
literature. \ldots\ It is \ldots\ very hard in practice when trying
to explain why one theory can be said to be more useful than another
to avoid at least tacitly appealing to the idea that it captures how
things are better than its rival does.''

Perhaps this is one aspect of Rorty you disagree with. But I
wonder about your saying that quantum mechanics is the best
justification for Rorty's philosophy, as if quantum mechanics is
something you accept to be real, an ``intrinsic nature of
reality'', the very idea of which Rorty explicitly rejects.
\ehw

First, just a technical point.  The philosophies I am most attracted
to at present are those of James and Dewey and what James says about
F.~C.~S. Schiller (but I haven't read Schiller himself yet).  Rorty
has donned himself to be the spokesman of those guys---and I don't
mind that because he writes so nicely---but his writings also have a
good admixture of the postmodernist ideas (of Foucault, Derrida,
etc.)\ thrown into them to boot.  This business about science not
being more trustworthy or real than literary criticism presently
strikes me as going too far.

But to Callinicos' point---``It is \ldots\ very hard in practice when
trying to explain why one theory can be said to be more useful than
another to avoid at least tacitly appealing to the idea that it
captures how things are better than its rival does.''---I would just
reply, ``Darwinism.''  And then, if that didn't sink in, I'd say,
``Darwinism.''  The point is, from this conception, there is very
little to say beyond that.  Were elephants written into the
blueprints of the universe?  From the Darwinistic conception, they
were not.  Yet, the species fills a niche and has had a stability of
at least a few million years worth.  There is a sense in which an
elephant, like a theory, is a ``true'' component in a description of
the world.  But that ``trueness'' only has a finite lifetime, and is
largely a result of a conspiracy of things beyond its command
(selection pressures).  To put it another way, in contrast to
Callinicos, the elephant doesn't ``capture how things are better than
its rival does'' in any absolute sense---only in a transitory
sense---but that doesn't take away from the functional value of the
elephant today.  So too, I am trying to imagine with theories.

Now, to quantum mechanics.  You find something contradictory about my
liking both quantum mechanics and Rorty.  Here is the way I would put
it.  Presently at least, I am not inclined to accept quantum
mechanics ``to be real, an `intrinsic nature of reality','' except
insofar as, or to the extent that, it is a ``law of thought,'' much
like simple (Bayesian) probability theory.$^2$\footnote{
\bq\noindent
$^2$ R.~Schack and I made this point best, I believe, in the
conclusion to our paper ``Unknown Quantum States and Operations, a
Bayesian View,'' quant-ph/0404156~\cite{Fuchs04}.  Let me restate it
here for clarity:\medskip

Is there something in nature even when there are no observers or
agents about? At the practical level, it would seem hard to deny
this, and neither of the authors wish to be viewed as doing so. The
world persists without the observer---there is no doubt in either of
our minds about that.  But then, does that require that two of the
most celebrated elements (namely, quantum states and operations) in
quantum theory---our best, most all-encompassing scientific theory to
date---must be viewed as objective, agent-independent constructs?
There is no reason to do so, we say.  In fact, we think there is
everything to be gained from carefully delineating which part of the
structure of quantum theory is about the world and which part is
about the agent's interface with the world.

{}From this perspective, much---{\it but not all}---of quantum
mechanics is about disciplined uncertainty accounting, just as is
Bayesian probability theory in general.  Bernardo and Smith write
this of Bayesian theory,
\bq
What is the nature and scope of Bayesian Statistics ... ?

Bayesian Statistics offers a rationalist theory of personalistic
beliefs in contexts of uncertainty, with the central aim of
characterising how an individual should act in order to avoid certain
kinds of undesirable behavioural inconsistencies.  The theory
establishes that expected utility maximization provides the basis for
rational decision making and that Bayes' theorem provides the key to
the ways in which beliefs should fit together in the light of
changing evidence.  The goal, in effect, is to establish rules and
procedures for individuals concerned with disciplined uncertainty
accounting.  The theory is not descriptive, in the sense of claiming
to model actual behaviour.  Rather, it is prescriptive, in the sense
of saying ``if you wish to avoid the possibility of these undesirable
consequences you must act in the following way.
\eq
In fact, one might go further and say of quantum theory, that in
those cases where it is not just Bayesian probability theory full
stop, it is a theory of stimulation and response. The agent, through
the process of quantum measurement stimulates the world external to
himself.  The world, in return, stimulates a response in the agent
that is quantified by a change in his beliefs---i.e., by a change
from a prior to a posterior quantum state.  Somewhere in the
structure of those belief changes lies quantum theory's most direct
statement about what we believe of the world as it is without agents.
\eq
}  Instead, I view quantum mechanics to be the first {\it rigorous\/}
hint we have that there might actually be something to James's
vision.

I've already told you the history of this, haven't I?  I gave a talk
in 1999 at Cambridge on the quantum de Finetti
theorem~\cite{Caves02}, after which Matthew Donald came up to me and
bellowed, ``You're an American pragmatist!''  I didn't know what that
meant really, but I kept the thought in the back of my head; I
figured one day, I'd figure out what he meant.  As it goes, that
happened on July 21 of last year.  I came across this book of Martin
Gardner's of which one of the chapters was titled, ``Why I Am Not a
Pragmatist''~\cite{Gardner83}. (Part of the story is recorded on page
15 of my little samizdat in a note titled ``The Reality of
Wives''~\cite{FuchsSamizdat}.  You might read it for a little laugh.)
As I read it, it was like a flash of enlightenment. For every reason
Gardner gave for not being a pragmatist, I thought about quantum
mechanics and realized that indeed I was one.  Donald was right after
all; I am an American pragmatist. And my further study of pragmatism
has borne that out to a T.

My point of departure, unlike James's, was not abstract philosophy.
It was simply trying to make sense of quantum mechanics, where I
think the most reasonable and simplest conclusion one can draw from
the Kochen-Specker results~\cite{Appleby03} and the Bell inequality
violations is, as Asher Peres says, ``unperformed measurements have
no outcomes''~\cite{Peres78}. The measurement provokes the ``truth
value'' into existence; it doesn't exist beforehand.  Now, go off and
read about James's and Dewey's theory of truth and you'll find almost
exactly the same idea (just without the rigor of quantum mechanics).
And similarly with lots of other pieces of the philosophy.

So, I view quantum mechanics as the hint of something much deeper.
But the full story is not yet told.  That is, quantum mechanics
strikes me as being to our community what the Galapagos Islands
were to Darwin---just a hint of something bigger.

\bhw
You and Rorty I guess would agree that ``dreams of a final
theory'' will never be more than dreams. I guess that idea does
not worry me as much as it would some physicists, but it does seem
like a defeat. But perhaps that just says something of my
personality. How much of a role does personality play in one's
preferred philosophy?
\ehw

Your question is a good one and one I worry about a lot.  Where your
knee-jerk reaction is defeat, mine is one of unlimited possibilities
and newfound freedom.  On a similar issue, James put it like this:
\bq
The history of philosophy is to a great extent that of a certain
clash of human temperaments.  Undignified as such a treatment may
seem to some of my colleagues, I shall have to take account of this
clash and explain a good many of the divergencies of philosophies by
it.  Of whatever temperament a professional philosopher is, he tries,
when philosophizing, to sink the fact of his temperament. Temperament
is no conventionally recognized reason, so he urges impersonal
reasons only for his conclusions. Yet his temperament really gives
him a stronger bias than any of his more strictly objective premises.
It loads the evidence for him one way or the other, making a more
sentimental or more hard-hearted view of the universe, just as this
fact or that principle would.  He {\it trusts\/} his temperament.
Wanting a universe that suits it, he believes in any representation
of the universe that does suit it. He feels men of opposite temper to
be out of key with the world's character, and in his heart considers
them incompetent and `not in it,' in the philosophic business, even
though they may far excel him in dialectical ability.

Yet in the forum he can make no claim, on the bare ground of his
temperament, to superior discernment or authority.  There arises
thus a certain insincerity in our philosophic discussions:  the
potentest of all our premises is never mentioned.  I am sure it
would contribute to clearness if in these lectures we should break
this rule and mention it, and I accordingly feel free to do so.
\eq

But I think the disparity between our views is in better shape than
that.  I think you're only seeing the program ``physics is the
ability to win a bet'' as a defeat because---even if you don't know
it---you're working within a kind of Kantian mindset.  That the
universe is already formed and there; that there is [a ``thing in
itself'' in no way dependent upon us]. Anything that can't be said
about the universe is then most surely a loss or limitation.  But, I
think once you see that what the pragmatist is trying to get at is
not that, maybe your heart will change.  Physics as the ability to
win a bet will strike you as something immensely positive.  Physics
is like that because reality is still forming, and the Darwinistic
component (along with the ``non-detachedness'' of the observer in
quantum mechanics) indicates that it may be somewhat malleable.  From
that point of view, to have ``dreams of a final theory'' is almost
like admitting defeat.

\section{to Several Correspondents, 23 April -- 5 December 2002, ``Snowflakes''}

This morning, the family and I woke up to find some real snow coming
down!  It's wonderful. I'm just taken with it.  A little while ago,
in the middle of my writing the paragraph above, I explained to Emma
the old childhood thing of how no two snowflakes are the same. That's
a thought that has been capturing a lot of my attention lately:  The
world and the snowflake.

\subsection{The POVM as a function from raw data to meaning}

We generally write a POVM as an indexed set of operators, $E_d$. Here
is how I would denote the referents of those symbols.  The index $d$
should be taken to stand for the raw data that can enter our
attention when a quantum measurement is performed.  The whole object
$E_d$ should be construed as the ``meaning'' we propose to ascribe to
that piece of data when/if it comes to our attention.  It is
important here to recognize the logical distinction between these two
roles.  The symbol $d$ stands for something beyond our control,
something that enters into us from the world outside our head.  The
ascription of a particular value $d$ is not up to us, by definition.
The {\it function\/} $E_d$, however, is of a completely different
flavor. It is set by our history, by our education, by whatever
incidental factors that have led us to believe whatever it is that we
believe when we walk into the laboratory to elicit some data.  That
is to say, $E_d$ has much the character of a subjective probability
assignment.  It is a judgment.

I have tried to say this in various ways before.  Maybe the first
place in {\sl Quantum States: W.H.A.T.?}~\cite{FuchsSamizdat} is in
the note ``Note on Terminology,'' pages 49--50, or in more detail in
``Replies to a Conglomeration,'' page 92.  Maybe there are still
better shots at it, but I didn't look further.  (I guess I also give
another variation on the matter on page 42 of
quant-ph/0205039~\cite{Fuchs02b}.) You can have a look at those if
you think it'll help, but I think the paragraph above says it as well
as anything.

\subsection{POVMs and radical pluralism}

Now let me go into a bit of the metaphysics of this.  Here's a point
of view that I'm finding myself more and more attracted to lately.

I think it is safe to say that the following idea is pretty
commonplace in quantum mechanical practice.  Suppose I measure a
single POVM twice---maybe on the same system or two different
systems, I don't care---and just happen to get the same outcome in
both cases.  Namely, a single operator $E_d$.  The common idea, and
one I've held onto for years, is that there is an objective sense in
which those two events are identical copies of each other.  They are
like identical atoms \ldots\  or something like the spacetime
equivalent of atoms.  But now I think we have no warrant to think
that.  Rather, I would say the two outcomes are identical only
because we have (subjectively) chosen to ignore almost all of their
structure.

That is to say, I now count myself not so far from the opinion of
Ulfbeck and Bohr, when they write~\cite{Ulfbeck01}:
\bq
\noindent
  The click \ldots\  is seen to be an event entirely beyond law.  \ldots\  [I]t
   is a unique event that never repeats \ldots\  The uniqueness of the click,
   as an integral part of genuine fortuitousness, refers to the click in
   its entirety \ldots\ . [T]he very occurrence of laws governing the clicks
   is contingent on a lowered resolution.
\eq

For though I have made a logical distinction between the role of the
$d$'s and the $E_d$'s above, one should not forget the very
theory-ladenness of the set of possible $d$'s to begin with.  What I
think is going on here is that it takes (a lot of) theory to get us
to even recognize the raw data, much less ascribe it some meaning. In
Marcus Appleby's terms, all that stuff resides in the ``primitive
theory'' (or perhaps some extension of it), which is a level well
below quantum mechanics.  What quantum mechanics is about is a little
froth on the top of a much deeper sea.  Once that deeper sea is set,
then it makes sense to make a distinction between the inside and the
outside of the agent---i.e., the subjective and the objective---as we
did above. For even in this froth on the top of a deeper sea, we
still find things we cannot control once our basic beliefs---i.e.,
our theory---are set.

Without the potential $d$'s we could not even speak of the
possibility of experiment.  Yet like the cardinality of the set of
colors in the rainbow---Newton said seven, Aristotle said three or
four~\cite{Zajonc93}---a subjective judgment had to be made (within
the wide community) before we could get to that level.  If this is
so, then it should not strike us as so strange that the raw data $d$
in our quantum mechanical experience will ultimately be ascribed with
a meaning $E_d$ that is subjectively given.  (I expressed some of
this a little better in a note I wrote to David [Mermin] last month;
I'll place it below as a supplement.)  All quantum measurements
outcomes are unique and incomparable at the ontic level.  At least
that's the idea I'm toying with.

\subsection{to N.~D. Mermin, 25 September 2002, ``Ulfbeck and Bohr''}

I finally got a chance to read the Ulfbeck/Bohr
paper~\cite{Ulfbeck01}. I know I've complained about Niels Bohr's
lack of detail when asserting the origin of the quantum formalism,
but I think they force my complaints to a whole new level.

There is, however, one idea in the paper that I am inclined to keep
or, at least to me, seems worth trying to develop.  I say this
predominantly because of its William Jamesian feel.  Here it is,
deleting the words of theirs that I don't like or don't agree with,
\bq
\noindent
  The click with its onset is seen to be an event entirely beyond law.
   \ldots\  [I]t is a unique event that never repeats \ldots\  The uniqueness of
   the click, as an integral part of genuine fortuitousness, refers to
   the click in its entirety, with all the complexity required for a
   break-through onto the spacetime scene. \ldots\  [T]he very occurrence of
   laws governing the clicks is contingent on a lowered resolution.
\eq

You see, from the Jamesian viewpoint of ``radical pluralism,'' every
piece of the universe, every crumb of its existence, is a unique
entity unto itself.  Here's a little quote in that direction from his
essay ``Abstrationism and `Relativismus'\,''~\cite{James97}:
\bq
Let me give the name of `vicious abstractionism' to a way of using
concepts which may be thus described: We conceive a concrete
situation by singling out some salient or important feature in it,
and classing it under that; then, instead of adding to its previous
characters all the positive consequences which the new way of
conceiving it may bring, we proceed to use our concept privatively;
reducing the originally rich phenomenon to the naked suggestions of
that name abstractly taken, treating it as a case of `nothing but'
that concept, and acting as if all the other characters from out of
which the concept is abstracted were expunged. Abstraction,
functioning in this way, becomes a means of arrest far more than a
means of advance in thought. It mutilates things; it creates
difficulties and finds impossibilities; and more than half the
trouble that metaphysicians and logicians give themselves over the
paradoxes and dialectic puzzles of the universe may, I am convinced,
be traced to this relatively simple source. {\it The viciously
privative employment of abstract characters and class names\/} is, I
am persuaded, one of the great original sins of the rationalistic
mind.
\eq

I wish I could find a better quote than that---I have memories of
reading the idea expressed in much greater detail and so much more
eloquently---but this morning, try as I might, I can't find it.

So I'll end this little note with another note I wrote a few months
ago to Greg Comer---it carries the sentiment, if not the eloquence.
It's pasted below.  Maybe I should have titled the present email, ``A
Click Is But a Click Not: It Is So Much More.''  For the same holds
with ``clicks'' as with ``atoms.''

\subsection{to G.~L. Comer, 23 April 2002, ``Music in the Musician''}

I think we just have to get rid of this imagery that we are ``made''
of atoms.  Or none of us are ever going to make any progress in our
emotional lives {\it or\/} our physical understanding.  By my present
thinking, a much better imagery is this.  Take me and an old log: we
both float in water.  That is to say, we have that much in common.
But there are a heck of a lot more things that we do not have in
common.  For any two entities, we can always find some
characteristics they have in common, if {\it we\/} are willing to
ignore all the ways in which they are distinct.  And that, I think,
is the story of atoms.  The atomic picture has something to do with
what we all have in common.  (Or, maybe more potently, it has
something to do with what is common in the {\it part\/} of existence
that we have chosen not to ignore for the moment.)  But to see the
atomic picture shine through, we have to dim down all the things that
are unique in us. Who said the particular shape of that rock is not
important?  Who said the pain you are feeling is only epiphenomena?

Such a picture of what physics and chemistry is about is every bit as
consistent as the worldview Steven Weinberg~\cite{Weinberg92}, say,
would have us believe.  And I would say that it is more so; for it
gives us a power and a hope for control in our lives that his can't
imagine.

\section{to N.~D. Mermin and R. Schack, 12 August 2003, ``Me, Me, Me''}

Me, me, me; it's always about me! ---Yes.  But nonetheless it is
simply not solipsism.  Let me explain.

I guess I was actually fortunate today:  For the second time in a
month, I was called a solipsist by one of my friends.  (This time the
accuser was Howard Wiseman.)  On top of that, Asher Peres gave a talk
this morning that made me cringe, saying things like, ``When no one
performs a measurement, nothing happens [in the world].''  The
combination of these two bad experiences caused me to wander the
streets of Aarhus this afternoon in spite of the horrible heat.  I
suppose I needed to find a way to sweat the poisons from my body.

The fortune in this is that it caused me once again to strive for a
clearer and more consistent form of expression.  I want to try to
capture some of that in this note.  Mostly it is about not allowing
oneself to get hung up in someone else's (inconsistent) expectations
for what quantum theory ought to be.

In our 2000 opinion piece in Physics Today~\cite{Fuchs00}, Asher and
I wrote:
\bq
\noindent
   The thread common to all the nonstandard ``interpretations'' is the
   desire to create a new theory with features that correspond to some
   reality independent of our potential experiments. But, trying to
   fulfill a classical worldview by encumbering quantum mechanics with
   hidden variables, multiple worlds, consistency rules, or spontaneous
   collapse, without any improvement in its predictive power, only
   gives the illusion of a better understanding. Contrary to those
   desires, quantum theory does {\it not\/} describe physical reality.
   What it does is provide an algorithm for computing {\it
   probabilities\/} for the macroscopic events (``detector clicks'')
   that are the consequences of our experimental interventions. This
   strict definition of the scope of quantum theory is the only
   interpretation ever needed, whether by experimenters or theorists.
\eq
But that is misleading and trouble-making.  In the second to last
sentence---with the experience of three more years of thinking on
this subject---I so wish we had said something more to the tune:
\bq
\noindent
   What quantum theory does is provide a framework for structuring MY
   expectations for the consequences of MY interventions upon the
   external world.
\eq
At least that is what the formal structure is about.  There is no
``we,'' there is no ``our.''  At this level of consideration, quantum
theory has nothing to do with intersubjective agreement. (By the way,
I'm not fooling myself:  Of course we could not have said what I said
above without restructuring the whole article---it would have opened
a can of worms!  I just want to try to do the idea better justice
right now.)

Here it is:  Any single application of quantum {\it theory\/} is
about ME, only me.  It is about MY interventions, MY expectations for
their consequences, and MY reevaluations of MY old expectations in
the light of those consequences.  It is noncommittal beyond that.
This is not solipsism; it is simply a statement of the subject
matter.

Is there any contradiction in this?  I say no, but how do I get you
into a mindset so that you might say the same?  Maybe the best way to
do this is to run through a glossary of quantum terms as I did once
before \ldots\ but now with all the latest slant.

\begin{itemize}
\item
SYSTEM:  In talking about quantum measurement, I divide the world
into two parts---the part that is subject to (or an extension of) my
will, and the part that is beyond my control (at least in some
aspects).  The idea of a ``system'' pertains to a part beyond my
control.  It counts as the source of my surprises, and in that sense
obtains an existence of its own external to me.  (Point \#1 against
solipsism, but I will return for another.)

\item
POVM:  In the theory, this counts as an extension of my will.  It
counts as a freely chosen action on my part.  The whole concept of a
``measuring device'' as something distinct from me---I am now
thinking---just gets in the way.  It is a point that Pauli made, but
I am coming ever more to appreciate it.  A ``measuring device'' is
like a prosthetic hand; its conceptual role is for the purpose of
recovering from our natural incapacities and, thus, might as well be
thought of as part of ourselves proper.  I perform a POVM on a
system---captured mathematically by a set---and one of its elements
comes about as a consequence.

\item
QUANTUM STATE:  As usual, the catalog of MY expectations for the
consequences of MY actions (i.e., POVMs) \ldots\ but now with
absolute, utter emphasis on the MY.

\item
UNITARY READJUSTMENT:  I'm talking here about the readjustment
appearing in Eq.~(95) of my paper quant-ph/0205039~\cite{Fuchs02b}.
This, like a quantum state, also captures a belief or expectation.
Its purpose is to quantify the extent to which I feel the need to
deviate from Bayes' rule after learning the consequence of my action.
This is what takes account of the nonpassive nature of MY
interventions.

\item
QUANTUM DYNAMICS:  This is the unitary readjustment (or mixture of
decompositions and unitary readjustments) that I judge I ought to
apply if my action on the system is passive, i.e., if my POVM is the
singleton set.  It is how I readjust my expectations when I am
learning nothing.
\end{itemize}

Summing up the glossary, I would say quantum theory in its single
user implementation is about ME.  I act on the world and it reacts in
a way unpredictable to me beyond the expectations I build from MY
quantum state (about the system).

Why is this not solipsism?  Because quantum theory is not a theory of
everything.  It is not a statement of all that is and all that
happens; it is not a mirror image of nature.  It is about me and the
little part I play in the world, as gambled upon from my
perspective.$^3$\footnote{\bq\noindent $^3$ See footnote 2.\eq} But
just as I can use quantum theory for my purposes, you can use it for
yours. Thus, if I had not been seeking dramatic effect above, I
should have more properly said, ``Any single application of quantum
{\it theory\/} is about the ME who applies it.''  (Don't correct my
English.) When David Mermin is a practitioner of quantum theory, what
the theory does is provide a framework for structuring HIS
expectations for the consequences of HIS interventions upon HIS
external world. \ldots\ And that is Point \#2 against solipsism.

Recall the definition of solipsism I dredged up from the {\sl
Encyclopedia Britannica\/}:
\bq
\noindent
   in philosophy \ldots\ the extreme form of subjective idealism that
   denies that the human mind has any valid ground for believing in
   the existence of anything but itself.
\eq

It seems to me we have plenty of valid ground for believing in the
existence of something besides ourselves:  It comes from all the
things we cannot control.  Indeed, as already emphasized, for those
things we can control, we might as well think of them as extensions
of ourselves.  Thus, to my mind, quantum theory already gives a
karate chop to solipsism because of the indeterminism it entails:
With each quantum measurement there is immediately something beyond
my control.

Beyond Point \#1, though, there is Point \#2.  It is a question of
finally getting straight what should and should not be in the purview
of the theory.  In this account, quantum theory is a theory of
personal action (and reaction).  The law-of-thought aspect of it
comes out with respect to each individual who uses it.  The textbook
poses an exercise that starts out, ``Suppose a hydrogen atom is in
its ground state.  Calculate the expectation of \ldots\ blah, blah,
blah.'' One might think it is asking us to calculate some objective
feature of the world.  It is not.  It is only asking us to carry out
the logical consequences of a supposed state of belief and a supposed
action that one could take upon the system.  And here's the clincher
about Bayesianism.  Just as no student in his right mind would find
it worthy to ask why the textbook writer posed the problem with the
ground state rather than the first excited state, no quantum theorist
should make a big to-do about it either.  It is simply an assumed
starting point.  An agent in the thick middle of a quantum
application can no more ask where he got his initial beliefs from,
than a pendulum can ask where it got its initial conditions from. The
cause of bottom-level initial conditions is {\it always\/} left
unanalyzed. If such was not a sin in Newtonian mechanics, it should
not be a sin in a Bayesian formulation of quantum mechanics.

So, it seems to me, if anything, the Bayesian account of quantum
theory is essentially the opposite of solipsism.  Rather than a unity
to nature, it suggests a plurality.  An image that might be useful
(but certainly flawed) comes from Escher's various paintings of
impossible objects.  The viewer would initially like to think of them
as two-D projections of a three-dimensional object; but he cannot.
Now imagine how much worse it would get if we were to have two
viewers with two slightly different paintings, each purporting to be
a different perspective on ``the'' impossible object.  Since neither
viewer can lift from his own two-D object to a three-D one, there is
no way to unify the pictures into a single whole.

Yet we live in one world, you say.  Maybe.  But, you should remember
that these quantum states we speak of are not perspectives.  They are
personal possessions.  To modify Tilgher's quote at the beginning of
de Finetti's ``Probabilismo''~\cite{DeFinetti31} for our own
purposes,
\bq
\noindent
   A quantum state is not a mirror in which a reality external to us
   is faithfully reflected; it is simply a biological function, a
   means of orientation in life, of preserving and enriching it, of
   enabling and facilitating action, of taking account of reality
   and dominating it.
\eq

``Are there other minds beside your own?,'' Howard Wiseman asks.  If
a mind is what it takes to write down a quantum state, then why not?
``If you leave the origin of the quantum state unanalyzed, why would
two minds ever agree on anything?''  That is the issue of
intersubjective agreement---something thankfully we can study within
the context of quantum theory.  But the first thing to get straight
is why the single user of quantum theory uses the very structure.
What is it precisely that he is believing of the world and his place
in it that leads him to the choice of quantum theory as his law of
thought?

That is, it is about ME and what I believe.  But what do I believe?
That's the research program!

\section{to A. Sudbery and H. Barnum, 18 August 2003, ``The Big IF''}

I have been trying to give Mr.\ Nagel a concerted effort during my
vacation here in Munich.  I went out and bought {\sl The View from
Nowhere}~\cite{Nagel86} and am a little way into it.

It's probably too early in my reading to tell, but my troubles with
Nagel may all boil down to ``The Big IF.''  That is, they may boil
down to the religion that lies behind this passage plucked out of his
article ``Subjective and Objective.''  (I'll capitalize the big IF
and a couple of other appropriate words so that you'll know what I'm
talking about.)  Here goes:
\bq
Since a kind of intersubjective agreement characterizes even what is
most subjective, the transition to a more objective viewpoint is not
accomplished merely through intersubjective agreement. Nor does it
proceed by an increase of imaginative scope that provides access to
many subjective points of view other than one's own. Its essential
character, in all the examples cited, is externality or DETACHMENT.
The attempt is made to view the world not from a place within it, or
from the vantage point of a special type of life and awareness, but
from nowhere in particular and no form of life in particular at all.
The object is to discount for the features of our pre-reflective
outlook that make things appear to us as they do, and thereby to
reach an understanding of things as they really are. We flee the
subjective under the pressure of an assumption that everything must
be something not to any point of view, but in itself. To grasp this
by DETACHING more and more from our own point of view is the
unreachable ideal at which the pursuit of objectivity aims.

Some version of this polarity can be found in relation to most
subject matter---ethical, epistemological, metaphysical. The relative
subjectivity or objectivity of different appearances is a matter of
degree, but the same pressures toward a more external viewpoint are
to be found everywhere. It is recognized that one's own point of view
can be distorted as a result of contingencies of one's makeup or
situation. To compensate for these distortions it is necessary either
to reduce dependence on those forms of perception or judgment in
which they are most marked, or to analyze the mechanisms of
distortion and discount for them explicitly. The subjective comes to
be defined by contrast with this development of objectivity.

Problems arise because the same individual is the occupant of both
viewpoints. In trying to understand and discount for the distorting
influences of his specific nature he must rely on certain aspects of
his nature which he deems less prone to such influence. He examines
himself and his interactions with the world, using a specially
selected part of himself for the purpose. That part may subsequently
be scrutinized in turn, and there may be no end to the process. But
obviously the selection of trustworthy subparts presents a problem.

The selection of what to rely on is based partly on the idea that the
less an appearance depends on contingencies of this particular self,
the more it is capable of being arrived at from a variety of points
of view. IF THERE IS A WAY THINGS REALLY ARE, which explains their
diverse appearances to differently constituted and situated
observers, then it is most accurately apprehended by methods not
specific to particular types of observers. That is why scientific
measurement interposes between us and the world instruments whose
interactions with the world are of a kind that could be detected by a
creature not sharing the human senses. Objectivity requires not only
a departure from one's individual viewpoint, but also, so far as
possible, departure from a specifically human or even mammalian
viewpoint. The idea is that if one can still maintain some view when
one relies less and less on what is specific to one's position or
form, it will be truer to reality. The respects in which the results
of various viewpoints are incompatible with each other represent
distortions of the way matters really are. And if there is such a
thing as the correct view, it is certainly not going to be the
unedited view from wherever one happens to be in the world. It must
be a view that includes oneself, with all one's contingencies of
constitution and circumstance, among the things viewed, without
according it any special centrality. And it must accord the same
DETACHED treatment to the type of which one is an instance. The true
view of things can no more be the way they naturally appear to human
beings than the way they look from here.

The pursuit of objectivity therefore involves a transcendence of the
self, in two ways: a transcendence of particularity and a
transcendence of one's type. It must be distinguished from a
different kind of transcendence by which one enters imaginatively
into other subjective points of view, and tries to see how things
appear from other specific standpoints. Objective transcendence aims
at a representation of what is external to each specific point of
view: what is there or what is of value in itself, rather than {\it
for\/} anyone. Though it employs whatever point of view is available
as the representational vehicle---humans typically use visual
diagrams and notation in thinking about physics---the aim is to
represent how things are, not {\it for\/} anyone or any type of
being. And the enterprise assumes that what is represented is
DETACHABLE from the mode of representation, so that the same laws of
physics could be represented by creatures sharing none of our sensory
modalities.
\eq

The two key ideas in this passage that I think quantum mechanics
plays the most havoc with are:
\begin{enumerate}
\item
the DETACHED agent (observer, scientist, etc.), and
\item
IF THERE IS A WAY THINGS REALLY ARE \ldots
\end{enumerate}

I honestly believe one can take the Nagel worldview seriously---I
suspect there is no logical flaw in it.  One can legitimately try to
make quantum mechanics fit that worldview with more or less success.
My only point is the strong personal suspicion that with such a
project one forces quantum mechanics into shoes it does not fit. And,
as I see it, what bunions that will cause in the future!

The whole subject matter of my {\sl Notes on a Paulian
Idea}~\cite{Fuchs03} is in toying with the idea that the cleanest
expression of quantum mechanics will come about once one realizes
that its overwhelming message is that the observer cannot be detached
from the phenomena he {\it helps\/} bring about.  I italicize the
word {\it helps\/} because I want you to take it seriously; the world
is not solely a social construction, or at least I cannot imagine it
so.  For my own part, I imagine the world as a seething orgy of
creation.  It was in that orgy before there were any agents to
practice quantum mechanics and will be in the same orgy long after
the Bush administration wipes the planet clean.  Both of you have
probably heard me joke of my view as the ``sexual interpretation of
quantum mechanics.''  There is no one way the world is because the
world is still in creation, still being hammered out. It is still in
birth and always will be---that's the idea.  What quantum mechanics
is about---I toy with---is each agent's little part in the creation
(as gambled upon from his own perspective).$^4$\footnote{\bq\noindent
$^4$ See footnote 2.\eq}  It is a theory about a very small part of
the world. In fact, I see it as a theory that is trying to tell us
that there is much, much more to the world than it can say.  I hear
it pleading, ``Please don't try to view me as a theory of everything;
you take away my creative power, my very promise, when you do that!
I am only a little theory of how to gamble in the light of a far more
interesting world!  Don't shut your eyes to it.''

The question is, how to get one's head around this idea and make it
precise?  And then, once it is precise, what new, wonderful, wild
conclusions can we draw from it?  That is the research program I am
trying to define.

Is it a {\it scientific\/} research program?  I think so, and in the
usual sense. There will be lemmas, theorems, and corollaries.  (I
would like to think that my work and the work of the fellows I've
drawn down this path already evidences this.)  Ultimately there will
be calls for experiments.  There will be technologies suggested and
money to be made from the program's fruits.  Failure of nerve?
Anything but!:
\bts
Maybe you and [Rorty] can shift me from my instinctive reaction to
pragmatism, which is that for a scientist it represents a failure of
nerve, a failure of imagination, and most seriously a failure of
curiosity. Being useful cannot, for a scientist, be the end of the
story about a statement or a theory; we immediately want to know
\underline{{\em why}} one theory is more useful than another. That
``why?''\ leads us to an external world of some kind, maybe very
strange (the stranger the better, i.e.\ the more interesting, I would
say) and to refuse to follow where it leads seems to me to be a
scientific copout.
\ets

I see it as anything but a failure of curiosity or a copout!  What
you wrote me above reminds me of a conversation I had with Chris
Timpson in a pub one night.  I made the mistake of mentioning William
James, and Chris quickly intoned, ``{\it All\/} James was about was
the nonsense that truth resides in what is useful.''  The word {\it
all\/} just boomed!  A man's whole life was dismissed in a single
sentence. I cut him short, ``William James was about many things,
{\it one of which\/} was that the correspondence theory of truth
holds no water.'' Similarly I will say to you, there is far more
explored by the pragmatist thinkers than that which is delimited by
their ideas on truth and warranted belief.  Pragmatism is not
positivism; it is not that there is nothing to be sought in science
beyond the connections between sense perceptions.  I see the
classical pragmatists (and myself) as ultimately realists, but honest
realists---ones who have realized that our theories are not mirror
images of the underlying reality, but rather extensions of our
biological brains.

But that is going in a direction I don't want to go down at the
moment.  In any case, don't read Rorty first!  Read James' little
book {\sl Pragmatism}~\cite{JamesPragmatism} to start off with.  More
immediately, with respect to the present Nagelian discussion, read
(in my {\sl Notes on a Paulian Idea}~\cite{Fuchs03}) ``Genesis and
the Quantum'' on pages 122--123, the dialogue between Adam and God on
pages 118--120, ``Evolution and Physics'' and ``Precision'' on pages
267--270, and some of Jeff Bub's expressions on the idea in Chapter
9, most notably pages 139--140 and 141--142---all these things in the
samizdat I sent you. The game of {\it assuming\/} the possibility of
a detached observer, as Nagel does, is just that:  a game of
assuming. Thereafter, Nagel tries to make sense of our more personal
worlds in spite of this. The pages I've just referred to in my
samizdat try to sketch what quantum mechanics might be talking about
if one does not make such an assumption.  In fact, they try to
justify {\it not\/} making the assumption at all.  I hope from these
readings you will get the impression that though there may be a
fundamental disagreement between Nagel and me at the outset, such a
disagreement does not necessarily amount to a copout on my part.

\section{to H.~Mabuchi, 17 June 2004, ``Preamble''}

I think I would like you to also post the little text file below
along with my other suggested readings for my ``Intro to QM''
lecture. You can give it the title ``Preamble''.  It was something I
sketched out on my flight over here, and reading over it again, I
kind of like it.

\bq
A lecturer faces a dilemma when teaching a course at a farsighted
summer school like this one.  This is because, when it comes to
research, there is often a fine line between what one thinks and what
is demonstrable fact.  More than that, conveying to the students what
one thinks---in other words, one's hopes, one's desires, the
potentest of one's premises---can be just as empowering to the
students' research lives (even if the ideas are not quite right) as
the bare tabulation of any amount of demonstrable fact.  So I want to
use one percent of this lecture to tell you what I think---the
potentest of all my premises---and use the remaining ninety-nine to
tell you about the mathematical structure from which that premise
arises.

I think the greatest lesson quantum theory holds for us is that when
two pieces of the world come together, they give birth.  [Bring two
fists together and then open them to imply an explosion.]  They give
birth to FACTS in a way not so unlike the romantic notion of
parenthood:  that a child is more than the sum total of her parents,
an entity unto herself with untold potential for reshaping the world.
Add a new piece to a puzzle---not to its beginning or end or edges,
but somewhere deep in its middle---and all the extant pieces must be
rejiggled or recut to make a new, but different, whole.  That is the
great lesson.

But quantum mechanics is only a glimpse into this profound feature of
nature; it is only a part of the story.  For its focus is exclusively
upon a very special case of this phenomenon:  The case where one
piece of the world is a highly-developed decision-making agent---an
experimentalist---and the other piece is some fraction of the world
that captures his attention or interest.

When an experimentalist reaches out and touches a quantum
system---the process usually called quantum `measurement'---that
process gives rise to a birth.  It gives rise to a little act of
creation.  And it is how those births or acts of creation impact the
agent's {\it expectations\/} for other such births that is the
subject matter of quantum theory.  That is to say, quantum theory is
a calculus for aiding us in our decisions and adjusting our
expectations in a QUANTUM WORLD\@.  Ultimately, as physicists, it is
the quantum world for which we would like to say as much as we can,
but that is not our starting point.  Quantum theory rests at a level
higher than that.

To put it starkly, quantum theory is just the start of our adventure.
The quantum world is still ahead of us.  So let us learn about
quantum theory.
\eq

\section{to G.~Musser, 7 July 2004, ``The Big Bang Is All Around Us''}

\bgm
[Fuchs said,] ``It is a theory about a very small part of the
world... a theory that is trying to tell us that there is much, much
more to the world than it can say.''  How is this not hidden
variables?

Sure, they may not be hidden variables in the pre-existing sense --
i.e.\ in the sense that a properly designed experiment can come
asymptotically close to ascertaining their pre-experiment value.  But
does not ``more to the world'' imply something hidden?
\egm

Take a break from me for a moment and ask yourself how the Everett
interpretation is not a hidden-variable theory?  (It almost seems you
would have asked the Everettian the same thing you asked me.)  A
hidden-variable theory is a very specific thing:  If one were to know
the value (even if only hypothetically and not operationally) of all
the variables (including possibly the ones on the inside of the
observer), then one can predict the outcome of all measurements with
certainty.  It is a fancy way of saying measurement outcomes
pre-exist, even if nothing one would ever call a measurement is
actually performed.

The determination or setting of specific measurement outcomes (in any
quantum mechanical experiment) has always been outside of the quantum
mechanical formalism.  There is nothing in the formalism that
determines whether one will get this click or whether one will get
that click in some measurement device.  But that does not make it a
hidden-variable theory.  What is hidden?

Here is the way Pauli put it~\cite{Pauli94}:
\bq
     Like an ultimate fact without any cause, the individual outcome
     of a measurement is, however, in general not comprehended by laws.
     This must necessarily be the case ...

     In the new pattern of thought we do not assume any longer the
     detached observer, occurring in the idealizations of this classical
     type of theory, but an observer who by his indeterminable effects
     creates a new situation, theoretically described as a new state of
     the observed system.  In this way every observation is a singling
     out of a particular factual result, here and now, from the
     theoretical possibilities, thereby making obvious the discontinuous
     aspect of the physical phenomena.

     Nevertheless, there remains still in the new kind of theory an
     objective reality, inasmuch as these theories deny any possibility
     for the observer to influence the results of a measurement, once
     the experimental arrangement is chosen.
\eq

(The conjunction of these thoughts is what I call ``the Paulian
idea''---hence the name of my book~\cite{Fuchs03}.)  ``Like an
ultimate fact without any cause, the individual outcome of a
measurement is not comprehended by laws.''

The way I see it, quantum measurement outcomes are ultimate facts
without specific call for further explanation.  And indeed the
quantum formalism supplies none.  Thus there is more to the world
than the quantum formalism can supply.  Nothing to do with hidden
variables.

But more specifically, regarding your point:
\bgm
``It is a theory about a very small part of the world... a theory
that is trying to tell us that there is much, much more to the world
than it can say.''  How is this not hidden variables?
\egm
How does the theory tell us that there is much more to the world than
it can say?  It tells us that {\it facts\/} can be made to come into
existence, and not just at some time in the remote past called the
``big bang'' but here and now, all the time, whenever an observer
sets out to perform (in antiquated language) a quantum measurement. I
find that fantastic!  And it hints that facts are being created all
the time all around us.  But that now steps out of the domain of what
the quantum formalism is about, and so is the subject of future
research.  At the present---as a first step---I want rather to make
the interpretation of the quantum formalism along these lines
absolutely airtight.  And then from there we'll better know how to go
further.

Doesn't that just make you tingle?  That (metaphorically, or maybe
not so metaphorically) the big bang is, in part, right here all
around us?  And that the actions we take are \underline{\it part\/}
of that creation! At least for me, it makes my life count in a way
that I didn't dare dream before I stumbled upon Wheeler, Pauli, and
Bell-Kochen-Specker.

But let me get away from this speculation and rope myself back in on
your particular question:  How is this not some hidden variables
account?  Simple:  If there are any extra facts being created around
us, they nevertheless do not impinge on the individual quantum
measurement outcome.

When I say that QM is a theory about a very small part of the world,
you should literally think of a map of the United States in relation
to the rest of the globe.  The map of the US is certainly incomplete
in the sense that it is obviously not a map of the whole globe.  But
on the other hand it is as complete as it can be (by definition) as a
representation of the US.  There are no hidden variables that one can
add to the US map that will magically turn into a map of the whole
globe after all.  The US map is what it is and need be nothing more.

Does that help any?

I think a good bit of the problem comes from something that was beat
into most of us at an early age.  It is this idea:  Whatever else it
is, quantum theory should be construed as a theory of the world.  The
formalism and the terms within the formalism somehow reflect what is
out there in the world.  Thus, if there is more to the world than
quantum theory holds out for, the theory must be incomplete.  And we
should seek to find what will complete it.

But my tack has been to say that that is a false image or a false
expectation.  Quantum theory from my view is not so much a law of
nature (as the usual view takes), but rather a law of thought.  In a
slogan:  Quantum mechanics is a law of thought.  It is a way of
plagiarizing George Boole who called probability theory a law of
thought.  (Look at the first couple of entries in the R\"udiger
Schack chapter of {\sl Notes on a Paulian Idea}~\cite{Fuchs03}.)  Try
to think of it in these terms, and let's see if this helps.

Let us take a simple term from probability theory, namely a
probability distribution over some hypothesis $P(h)$.  This function
represents a gambling agent's expectations about which value of $h$
will obtain in an observation or experiment.  Suppose now the agent
gathers a separate piece of data d from some other observation or
experiment and uses it to conditionalize his expectations for $h$;
i.e., he readjusts his expectations for h to some new function
$P(h|d)$ by using Bayes' rule.  Now here's a question for you.  Is
there anything within abstract probability theory that will allow the
agent to predict precisely which value of $d$ he will find when he
gathers his data?  Of course not.  It's almost silly to pose the
question. Abstract probability theory has nothing to do with the
actual facts of the world.  But then, doesn't that mean that
probability theory is an incomplete theory?  It can't, for instance,
explain its own transitions $P(h) \longrightarrow P(h|d)$ since
probability theory alone can't tell us why this $d$ rather than that
$d$. Moreover if probability is incomplete in this way, shouldn't we
be striving to complete it? Both silly questions, and I hope for
obvious reasons.

So:
\begin{enumerate}
\item
There is no particular mystery in the transition $P(h)
\longrightarrow P(h|d)$.
\item
We would never expect probability theory to provide a mechanism to
determine which value of $d$ is found or produced in the experiment.
The value $d$ represents a fact of the world, and probability theory
is {\it only\/} a theory about how to manipulate expectations once
facts are given.
\item
But also no one would be compelled to call probability theory
incomplete because of this.
\item
In particular, admitting this does not amount to having a
hidden-variable explanation of probability theory.
\end{enumerate}

So I say with quantum mechanics.  The story is almost one-to-one the
same:  You just replace probability distributions with quantum
states.  ... But then you reply, ``But there's a difference; quantum
theory is a theory of physics, it is not simply a calculus of
thought.''  And I say, ``That's where you err.''  Quantum theory
retains a trace of something about the real, physical world but
predominantly it is a law of thought that agents should use when
navigating in the (real, physical) world.  In particular, just like
with probability theory, we should not think of quantum theory as
incomplete in the usual sense.  If it is incomplete in any way, it is
only incomplete in the way that the US map is incomplete with respect
to the globe: There's a lot more land and ocean out there.

``Teasing out'' (your words) the trace of the physical world in the
formalism---i.e., the part of the theory that compels the rest of it
as a useful law of thought---is the only way I see to get a solid
handle on what quantum mechanics is trying to tell us about nature
itself.

With this let me now go back to the US map for one final analogy.  I
said that there is a sense in which the US map is as complete as it
can be.  However there is also a sense in which it tells us something
about the wider world:  If we tabulate the distances between cities,
we can't help but notice that the map is probably best drawn on the
surface of a globe.  I.e., the US already reveals a good guess on the
curvature of the world as a whole---it hints that the world is not
flat.  And that's a great addition to our knowledge!  For it tells a
would-be Columbus that he can safely go out and explore new
territories.  Exploring those new territories won't make the US map
any more complete, but it still means that there is a great adventure
in front of him.

\begin{theacknowledgments}

I thank Marcus Appleby, Howard Barnum, Harvey Brown, Jeff Bub, Carl
Caves, Greg Comer, Matthew Donald, Hideo Mabuchi, David Mermin,
George Musser, Steve Savitt, R\"udiger Schack, Tony Sudbery, Chris
Timpson, and Howard Wiseman for their questions, their protests, and
their ears.
\end{theacknowledgments}

\end{document}